\documentclass[draft]{agujournal}
\usepackage[T1]{fontenc}
\usepackage[latin9]{inputenc}
\usepackage{array}
\usepackage{multirow}
\usepackage{graphicx}
\usepackage{wasysym}
\usepackage[unicode=true]
 {hyperref}

\makeatletter

\providecommand{\tabularnewline}{\\}

\journalname{Space Weather}
\let\maketitle\relax
\let\bibliographystyle\relax

\makeatother

\begin{document}

\title{First results from sonification and exploratory citizen science of
magnetospheric ULF waves: Long-lasting decreasing-frequency poloidal
field line resonances following geomagnetic storms}
\maketitle

\authors{M. O. Archer \affil{1,2}, M. D. Hartinger \affil{3}, R. Redmon
\affil{4}, V. Angelopoulos \affil{5}, B. M. Walsh \affil{6}, and
Eltham Hill School Year 12 Physics students \affil{7}}

\affiliation{1}{School of Physics \& Astronomy, Queen Mary University of London,
London, UK.}

\affiliation{2}{Space \& Atmospheric Physics Group, Blackett Laboratory, Imperial
College London, London, UK.}

\affiliation{3}{Electrical and Computer Engineering Department, Virginia Tech, Blacksburg,
VA, USA.}

\affiliation{4}{National Oceanic and Atmospheric Administration, Boulder, CO, USA.}

\affiliation{5}{Department of Earth, Planetary and Space Sciences, University of
California, Los Angeles, CA, USA.}

\affiliation{6}{Department of Mechanical Engineering and Center for Space Physics,
Boston University, Boston MA, USA.}

\affiliation{7}{Eltham Hill School, London, UK.}

\correspondingauthor{M. O. Archer}{ m.archer10@imperial.ac.uk}
\begin{keypoints}
\item Sonification can be used for exploratory citizen science, in
this case of ultra-low frequency waves at geostationary orbit\\
\item Long-lasting narrowband poloidal field line resonances with
decreasing frequency have been identified\\
\item Such events occur more often than previously thought, typically
in the recovery phase of geomagnetic storms
\end{keypoints}
\begin{abstract}
Magnetospheric ultra-low frequency (ULF) waves contribute to space
weather in the solar wind \textendash{} magnetosphere \textendash{}
ionosphere system. The monitoring of these waves by space- and ground-based
instruments, however, produces ``big data'' which is difficult to
navigate, mine and analyse effectively. We present sonification, the
process of converting an oscillatory time-series into audible sound,
and citizen science, where members of the public contribute to scientific
investigations, as a means to potentially help tackle these issues.
Magnetometer data in the ULF range at geostationary orbit has been
sonified and released to London high schools as part of exploratory
projects. While this approach reduces the overall likelihood of useful
results from any particular group of citizen scientists compared to
typical citizen science projects, it promotes independent learning
and problem solving by all participants and can result in a small
number of unexpected research outcomes. We present one such example,
a case study identified by a group of students of decreasing-frequency
poloidal field line resonances over multiple days found to occur during
the recovery phase of a CME-driven geomagnetic storm. Simultaneous
plasma density measurements show that the decreasing frequencies were
due to the refilling of the plasmasphere following the storm. The
waves were likely generated by internal plasma processes. Further
exploration of the audio revealed many similar events following other
major storms, thus they are much more common than previously thought.
We therefore highlight the potential of sonification and exploratory
citizen science in addressing some of the challenges facing ULF wave
research.
\end{abstract}
\textbf{Plain Language Summary} Earth\textquoteright s magnetic shield,
protecting us against harmful radiation from the Sun and more distant
sources, is rife with a symphony of ultra-low frequency analogues
to sound waves. These waves transfer energy from outside this shield
to regions inside it and therefore play a key role in space weather
- how space poses a risk to our everyday lives by affecting power
grids, GPS, passenger airlines, mobile telephones etc. While these
waves are too low pitch for us to hear them, we can make our satellite
recordings of them audible by dramatically speeding up their playback.
We show that these audio versions of the data can be used by school
students to contribute to research, by having them explore the data
through the act of listening and performing analysis using audio software.
An example of this is presented where London school students identified
waves whose pitch decreased over the course of several days. The waves
were the natural oscillations of magnetic field lines, like the vibrations
of a plucked guitar string which forms a distinct note. The changing
pitch is explained by the fact that the event occurred as Earth's
magnetic shield was recovering from a disturbance caused by a ``solar
storm''. Many similar events were discovered in the audio which also
followed such disturbances, revealing that these types of waves are
much more common than previously thought. Therefore we have demonstrated
that making data audible and involving the public can further the
research into space weather.

\textbf{Index Terms and Keywords }\\
2752 MHD waves and instabilities \\
2788 Magnetic storms and substorms (4305, 7954)\\
9820 Techniques applicable in three or more fields 

\section{Introduction}

Ultra-low frequency (ULF) magnetohydrodynamic waves, with periods
between seconds and tens of minutes, transfer mass, energy and momentum
throughout the solar wind \textendash{} magnetosphere \textendash{}
ionosphere system. They can accelerate auroral and radiation belt
electrons via wave-particle interactions and resonances \citep{elkington13,chaston13},
and routinely make significant contributions to local (and in extreme
cases global) Joule-heating of the ionosphere/thermosphere \citep{hartinger15}.
Many different modes of ULF waves may be excited within Earth's magnetosphere
driven by a variety of both internal and external processes whose
properties contain information about the processes that generated
them and the regions through which they have propagated, resulting
in a \textquotedblleft zoo\textquotedblright{} of different ULF wave
phenomena \citep{mcpherron05,wright06}. One example mode is the field
line resonance (FLR), standing Alfv\'{e}n waves on field lines approximately
fixed at conjugate ionospheres \citep{southwood74}. It is known that
FLRs' occurrence and properties (such as frequency) can be highly
variable depending on the solar wind and magnetospheric driving conditions
present \citep[e.g.][]{takahashi10,takahashi14,archer15b,archer17}.
However, historically ULF waves in general have merely been classified
as either quasi-sinusoidal or irregular and split into frequency bands,
where the limits of these bands are not precise and more than one
process may produce waves in a particular (or across multiple) band(s)
\citep{mcpherron05}. Statistical studies often only consider how
the integrated power over one or many of these bands varies with conditions
\citep[e.g.][]{mann04}, hence narrowband or multi-harmonic oscillations
are not distinguished from broadband features in such studies. In
cases where the former are considered, the methods used are often
manual or semi-manual \citep[e.g.][]{takahashi15}. Given the vast
amount of ULF wave data being produced by both space- and ground-based
instruments, new techniques could potentially help.

Sonification is the use of non-speech audio to convey information
or perceptualize data \citep{kramer94}. Using the human auditory
system has several advantages in temporal, spatial, amplitude, and
frequency resolution over visualization techniques. For example the
human hearing range of 20\textendash 20,000~Hz spans three orders
of magnitude in frequency and at least 13 orders of magnitude in sound
pressure level \citep{robinson56}, whereas the human visual system's
perceptible frequencies range over only a quarter of an order of magnitude
and no more than 4 orders of magnitude in luminance \citep{kunkel10}.
Human hearing is highly nonlinear and has been shown to identify the
pitch and timing of sound signals much more precisely than allowed
by the Gabor limit or uncertainty principle \citep{oppenheim13},
which is a consequence of linear analysis methods such as the Fourier
or wavelet transforms. While nonlinear analysis methods exist such
as the Wigner-Ville distribution \citep{wigner32,ville48} or Empirical
Mode Decomposition \citep{huang98}, these often introduce artefacts,
mode mixing or can be unstable. Applications to magnetospheric ULF
waves have shown that linear and nonlinear methods all have their
own advantages and drawbacks depending on the precise nature of the
waves present \citep{chi08,piersanti18}. Furthermore, the human auditory
system's ability to separate sounds corresponding to different sources
far outperforms even some of the most sophisticated blind source separation
algorithms developed to date \citep{divenyi05}. Therefore, it is
clear that there should also be a place for data sonification in addition
to both standard and more recent methods of analysis.

There is a long history of converting space plasma physics data into
audible sounds, for example the terminology of ionospheric extremely-low
frequency (ELF) and very-low frequency (VLF) waves, which largely
span the human hearing range, such as ``whistlers'' \citep{barkhausen19}
and ``lion roars'' \citep{smith67} were based on their psychoacoustics
when picked up by radio antenna. This tradition has continued with
nomenclature such as ``tweaks'', ``chorus'', ``hiss'' and ``static''
being commonly used in magnetospheric and ionospheric ELF/VLF research
and various ELF/VLF datasets from across the solar system are available
in audio format (e.g. \href{http://space-audio.org/}{http://space-audio.org/}).

Sonification techniques can also be used for waves normally outside
of the human auditory range. Depending on the time resolution of the
data and the frequency ranges of interest, sonification can dramatically
cut down on the analysis processing time making it ideal for effective
navigation, mining and analysis of ``big data'' within exploratory
research \citep{hermann2002}. \citet{alexander11,alexander14} and
\citet{wicks16} showed that researchers using sonified solar wind
data were able to identify subtle features embedded within the data
that were not necessarily clear from standard visual analysis techniques.
However, sonification of magnetospheric ULF wave datasets is not widespread.
The only publicly available example we are aware of is the CARISMA
ground magnetometer network's induction coil magnetometer data (\href{http://www.carisma.ca/}{http://www.carisma.ca/}).
Each day of 20~Hz resolution data is filtered and resampled as 44,100~Hz
audio, allowing waves of frequencies down to $9\,\mathrm{mHz}$ in
theory (though in practise the lower limit will likely be higher depending
on the subject) to be heard. However, this entirely neglects the Pc5
band of ULF waves (2\textendash 7~mHz) and some of the Pc4 band also
(7\textendash 22~mHz), both of which play important roles within
the magnetosphere.

Another fairly recent technique to tackle large datasets is that of
citizen science, involving \textquotedblleft organized research in
which members of the public engage in the processes of scientific
investigations by asking questions, collecting data, and/or interpreting
results\textquotedblright{} (\href{http://www.citizenscience.org}{CitizenScience.org}).
It typically works by the concept of crowdsourcing and can collect
data and/or analysis which may be extremely difficult and/or inefficient
to be carried out either by a small number of researchers or by using
computational algorithms. This crowdsourcing model therefore requires
well thought out, highly prescribed activities to be undertaken by
all the citizen scientists. It should also be noted that as well as
the benefits to the research, citizen science should also positively
impact upon the volunteers who participate through their educational
and outreach/engagement elements. While formalised citizen science
projects are well established in areas such as astronomy and biology,
they are rare in space plasma physics \citep{knipp15}. Solar Stormwatch,
integrated into the \href{http://www.zooniverse.org/}{Zooniverse.org}
platform, uses citizen scientists to track the propagation of Coronal
Mass Ejections (CMEs) through the heliosphere \citep{barnard14}.
Sunspotter (\href{http://www.sunspotter.org}{http://www.sunspotter.org})
tasked citizen scientists to rank images of sunspots by their complexity.
Aurorasaurus maps the location of the auroral oval by taking advantage
of geotagged posts on Twitter which are verified by citizen scientists
as having been auroral observations \citep{macdonald15}. Clearly
there is scope to do more in this area within the field.

This paper presents a project which uses sonification to enable exploratory
citizen science research into magnetospheric ULF waves called Magnetospheric
Undulations Sonified Incorporating Citizen Scientists (MUSICS, \href{http://www.qmul.ac.uk/spa/musics}{http://www.qmul.ac.uk/spa/musics}).
Magnetometer data at geostationary orbit was made audible in the ULF
range and given to high school students as part of independent projects.
We present the sonification process, framework of the exploratory
citizen science, and first-results emerging from the MUSICS project.

\section{MUSICS project}

\subsection{Sonification}

Whilst many different forms of sonification are possible, we use the
simplest method which is sometimes known as audification. This is
the direct translation of time-series data to audio samples, hence
is only applicable to an already oscillating signal and thus clearly
relevant to the study of magnetospheric ULF waves. The sonified data
can then be analysed by the human auditory system, rather than the
forms of visual analysis usually performed. \citet{alexander14} provide
a detailed list of considerations when attempting this process, here
we detail the methods chosen for use in ULF wave sonification applied
to Geostationary Environment Operational Satellite (GOES) data.

The sonification was applied to each year of GOES magnetometer data
at 512~ms resolution. Currently this data is available for 2007\textendash 2008
(GOES-10, -11 and -12) as well as 2010\textendash 2017 (GOES-13 and
-15). Before sonification, the ULF waves must first be extracted from
the data and transformed into an appropriate coordinate system - a
fairly standard procedure. The original data in PEN coordinates were
used, where P is perpendicular to the satellite\textquoteright s orbital
plane, E lies parallel to the satellite\textendash Earth center line
and points earthward, and N is perpendicular to both pointing eastwards.
A mean-field aligned coordinate system was defined by taking a 34~min
running average of the data, whose direction at each time shall henceforth
be called the compressional component (com) since magnetic field oscillations
in this direction are representative of compressional modes. The two
remaining directions defining the coordinate system are the poloidal
component (pol), corresponding to the direction perpendicular to the
mean field pointing radially outwards from the Earth; and the toroidal
component (tor), perpendicular to both and directed azimuthally towards
east. The data was transformed into this coordinate system and the
running average subtracted from the compressional component, thereby
extracting ULF waves of frequencies $\apprge0.5\,\mathrm{mHz}$. Any
data gaps were interpolated for the coordinate transformation and
subsequently filled with zeros to ensure a full year of regularly
sampled data for the sonification.

The choice of a lowest frequency scale of $0.5\,\mathrm{mHz}$ was
made for numerous reasons. The first concerns known and/or theorised
ULF wave modes. The magnetospheric density survey of \citet{archer15b}
showed that the lowest expected fundamental frequencies of FLRs at
geostationary orbit in the dawn, noon, and dusk sectors across half
a solar cycle were all around $0.7\,\mathrm{mHz}$. Furthermore, the
proposed eigenmode of the dayside magnetopause is also expected to
typically have similarly low frequencies, with these being less than
$0.5\,\mathrm{mHz}$ only 10\% of the time \citep{archer15a}. The
second reason for the choice of scale pertains to spacecraft motion.
Over $34\,\mathrm{min}$ the GOES spacecraft's azimuthal position
changes by about $8^{\circ}$, however, over much longer time scales the
spacecraft motion can no longer be neglected and one risks mixing
spatial and temporal effects \citep{urban16}. Furthermore, when considering
such long timescales the mean field no longer becomes representative
of the background geomagnetic field, rendering the field-aligned coordinate
system inadequate at distinguishing between the physical processes
behind the ULF waves.

For effective conversion to audio, it is necessary to tailor the sampling
rate of the output such that frequencies of interest will map to the
human auditory range of approximately 20\textendash 20,000~Hz. The
relationship between the real frequency of a signal in the data $f_{real}$
to that in the audio $f_{audio}$ is given by

\begin{equation}
f_{audio}=f_{real}\times F_{s}\Delta t_{real}
\end{equation}
where $\Delta t_{real}$ is the cadence of the original data and $F_{s}$
the sampling frequency of the outputted audio. A widespread standard
sampling frequency in audio is $F_{s}=$44,100~Hz since this this
corresponds to a Nyquist frequency just above the upper threshold
for humans. However, this simple mapping would render the lowest frequencies
present in the data inaudible. The data was therefore boxcar-averaged
by 4 datapoints, downsampling its resolution to $\Delta t_{real}=2.048\,\mathrm{s}$
before the conversion. The sonification, therefore, has made the vast
majority of all physically meaningful frequencies in the GOES data
audible.

Two different types of audio were produced: a straightforward conversion
of the magnetic field data; and a spectrally whitened dataset produced
by first time-differencing the data \citep[e.g.][]{takahashi10} and
then sonifying. Since audio waveforms are unitless and must lie within
the range -1 to 1, some form of amplitude scaling must also be performed.
In order to preserve the natural variability of the data, we simply
scale the data by dividing through by factors of $10~\mathrm{nT}$
and $0.49~\mathrm{nT}~\mathrm{s}^{-1}$ respectively - less than typical
ULF wave amplitudes \citep[e.g.][]{takahashi12}. Any values greater
than these thresholds were clipped to the maximum waveform values
of 1 or -1. Clipping occurred less than 3\% of the time.

While each component was output separately as a mono audio file, a
summary stereo file was also produced corresponding to the poloidal
component in the left channel, the toroidal component in the right
channel and the compressional channel multiplied by a factor of 0.5
added to both channels. This enables one to listen to all three components
at once and perform a cursory polarisation analysis (particularly
when using headphones) by judging whether any identified events are
loudest in either the left or right ears or approximately of equal
loudness in both ears.

Through this sonification process each year of data is converted into
approximately 6~min of audio. The time within the audio is given
by
\begin{equation}
t_{audio}=\frac{t_{real}}{F_{s}\triangle t_{real}}\label{eq:time}
\end{equation}
assuming that the actual time $t_{real}$ starts at zero at the beginning
of each year. Ogg Vorbis compression was used to write the audio files
since this does not suffer the issues of MP3 encoders, which introduce
silence at the beginning of audio files thereby rendering the time
conversions in equation~\ref{eq:time} incorrect. Given the longitudes
of the GOES spacecrafts' geostationary orbits, local time can also
be easily calculated. Therefore, the sonification dramatically reduces
the timeframe of the data, allowing for fast surveying of ULF wave
``big data'' with the ability to ascertain the real times and locations
of events.

\subsection{Citizen Science}

While sonified data has been used as a tool for researchers, we are
not aware of its widespread use in exploratory citizen science projects.
However, given the ease of data navigation, mining and analysis afforded
we believed it would lend itself well to such efforts by lowering
the barrier to entry of getting involved. Furthermore, since everyone's
perception of sound is different and pattern recognition (particularly
of weak or noisy signals) within audio improves with practice \citep{whitton14},
utilising a wide pool of citizen scientists listening to the same
dataset should identify numerous different types of events, whereas
use of sonified data by a single researcher might only highlight just
one or a few. Therefore a project was developed for London schools
as part of a wider initiative enabling high school students to experience
research science and independently develop their own investigations.
The main aim of this initiative is on raising students' aspirations
towards Physics (or STEM), with potential benefits towards the research
being of secondary concern. This is somewhat different to most citizen
science, where addressing the scientific objectives are of primary
importance. To highlight this difference in aim as well as the educational
environment in which they are set, such initiatives are sometimes
called 'Research in Schools' rather than citizen science despite the
similarities between the two approaches. Details on the first couple
of years of the entire 'Research in Schools' programme, preliminary
results on the impact it has had on students and teachers, and the
lessons learned about setting up such a programme can be found in
\citet{archer_report17}. To summarise, students report having increased
their confidence in various topics and scientific methods related
to their project area to a high level of statistical significance
(currently the MUSICS project has an overall $6.1$ z-score in a Wilcoxon
signed-rank test \citep{wilcoxon45}), as well as developing a wide
array of different skills many of which they would likely not have
had access to previously in their school environment. Teachers fed
back having learnt new physics content and developed skills which
could be implemented or referred to in future lessons, as well as
gaining confidence in supporting their students and discussing research
content with them.

\begin{figure}
\includegraphics[width=1\columnwidth]{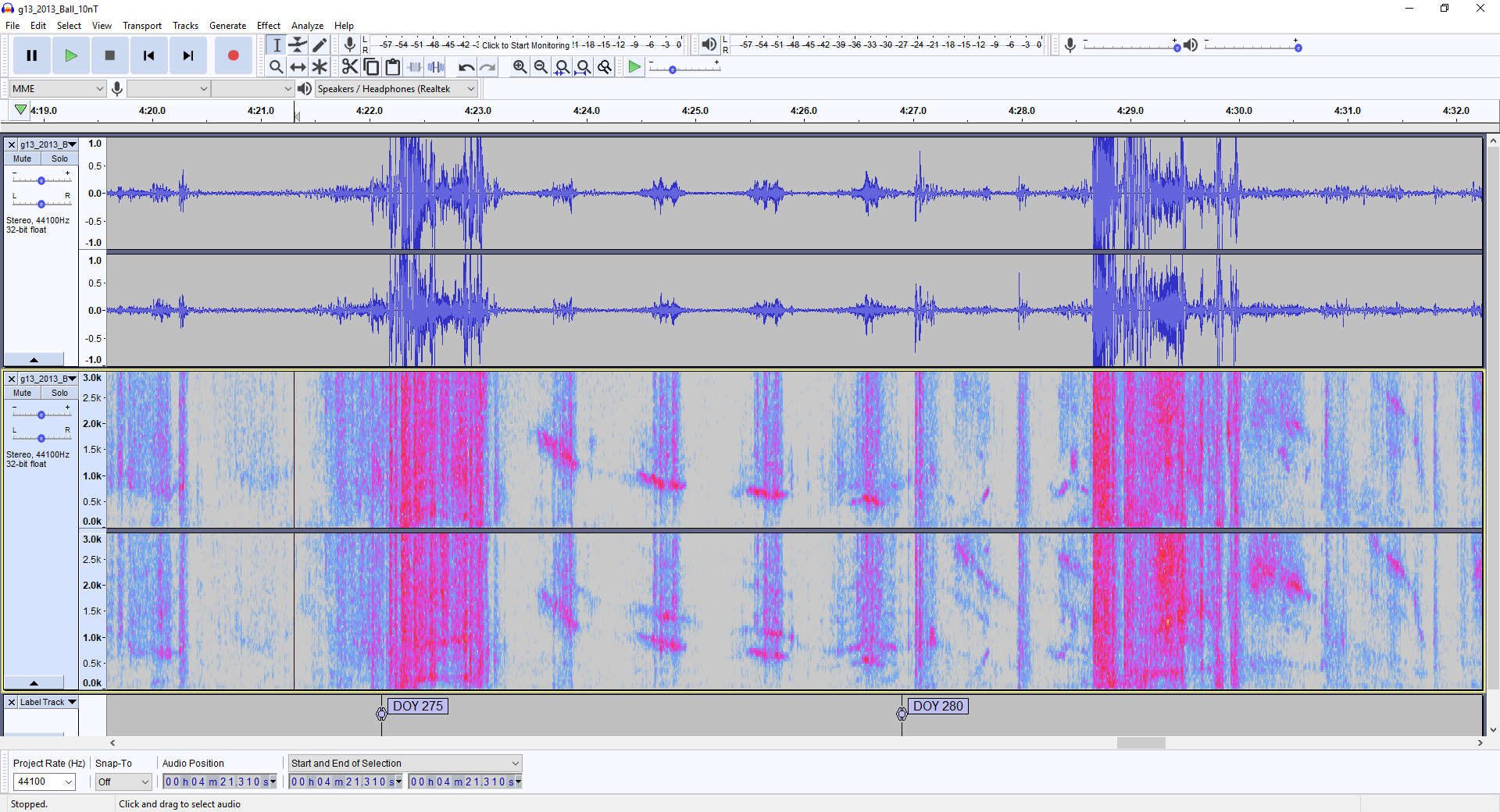}

\caption{Citizen scientists' view in Audacity of the event presented in Section~\ref{sec:First-Results}.
The top tracks show the waveform view of the GOES-13 summary stereo
audio file whereas the bottom tracks show the spectrogram view of
the GOES-13 time-differenced summary stereo audio file. Labels have
been added in the software to denote day-of-year (DOY) ranges.\label{fig:audacity}}

\end{figure}

The MUSICS project runs for 6~months each year from around the start
of the UK autumn term and ending before the Easter/spring break, after
which the students involved (ranging from 14\textendash 18 years old)
typically have exams. At the start of the project the students are
introduced to the area of magnetospheric ULF wave research and are
given a written introductory guide, the sonified data, free Audacity
audio software (\href{http://www.audacityteam.org/}{http://www.audacityteam.org/})
and a standard spreadsheet for logging events. The audio software
allows the students to perform several standard analysis methods simply
within the graphical user interface including plotting power spectra,
spectrograms, correlation-based pitch analysis, root mean square amplitude
computations etc. with an example view shown in Figure~\ref{fig:audacity}.
Indeed much of the analysis presented in section~\ref{sec:First-Results}
are possible within the software requiring no programming expertise
by the students. The spreadsheet provided implements conversions of
time, local time, frequency and amplitude applied to the sonified
data. Students are encouraged to first explore the data and the tools
provided before in small groups focusing on investigating a single
or class of similar events that they have identified aurally. Throughout
the project they are supported both by their teacher and a number
of visits from researchers, providing suggestions of how they can
progress with their project as well as the wider context of what they've
done in terms of previous research. At the end of the project they,
along with students working on other projects within the wider programme,
present their work either as a talk or poster at a special student
conference.

Such an exploratory model of citizen science has its benefits and
drawbacks. Most citizen science is highly prescriptive to the participants,
since they are presented with one or a series of carefully constructed
questions or tasks to follow. In this respect the citizen scientists
are largely being used to crowdsource data/analysis which may be difficult
to implement by individuals or algorithmically. Therefore, the focus
is on producing a usable dataset rather than on the independent learning
and problem solving of the participants. Such a model has its benefits
as the aggregated citizen science data may easily be implemented in
tackling the intended research. Because of this, however, these projects
primarily address one or a limited number of science questions. While
there is sometimes scope for unexpected results, typically through
citizen scientists discussing and collaborating with researchers on
discussion boards, such activities are undertaken by a very small
minority of citizen scientists.In contrast, the 'Research in Schools'
style of projects, such as MUSICS,provide much less prescription than
standard citizen science because the focus is on independent learning
and problem solving by the participants, an important aspect of a
researcher's experience which typical citizen science tends to emphasize
less, in order to positively impact upon the students involved. All
citizen scientists in this model are encouraged to tailor or adapt
their methods depending on what they have discovered, through collaboration
with researchers. This is the primary methodology used for all involved,
rather than something undertaken by only a fraction of participants.
Of course, the overall likelihood of useful results from any particular
group of citizen scientists towards the research is reduced in this
case compared to typical citizen science. Nonetheless, in the next
section we show that this model of citizen science can indeed result
in unexpected research outcomes. We therefore stress that both models
of citizen science have their merits.

\section{First Results\label{sec:First-Results}}

To demonstrate the potential of using sonified ULF wave data in citizen
science projects for original research, the first results from the
MUSICS project are presented. Students at Eltham Hill School in South
East London aurally identified a case study event from the sonified
GOES magnetometer dataBased on the students' initial work, here we
present the results of more detailed research performed by professional
scientists into that event. Thus the scientific results presented
in this paper were made possible only thanks to the citizen scientists'
identification of the event and their preliminary analysis of it.
Note that the analysis here uses the GOES magnetometer data in the
mean-field aligned coordinate system but before its conversion into
audio, so no issues arise due to e.g. clipping of waveforms.

\subsection{ULF Wave Observations}

\begin{figure}
\begin{centering}
\includegraphics[width=1\columnwidth]{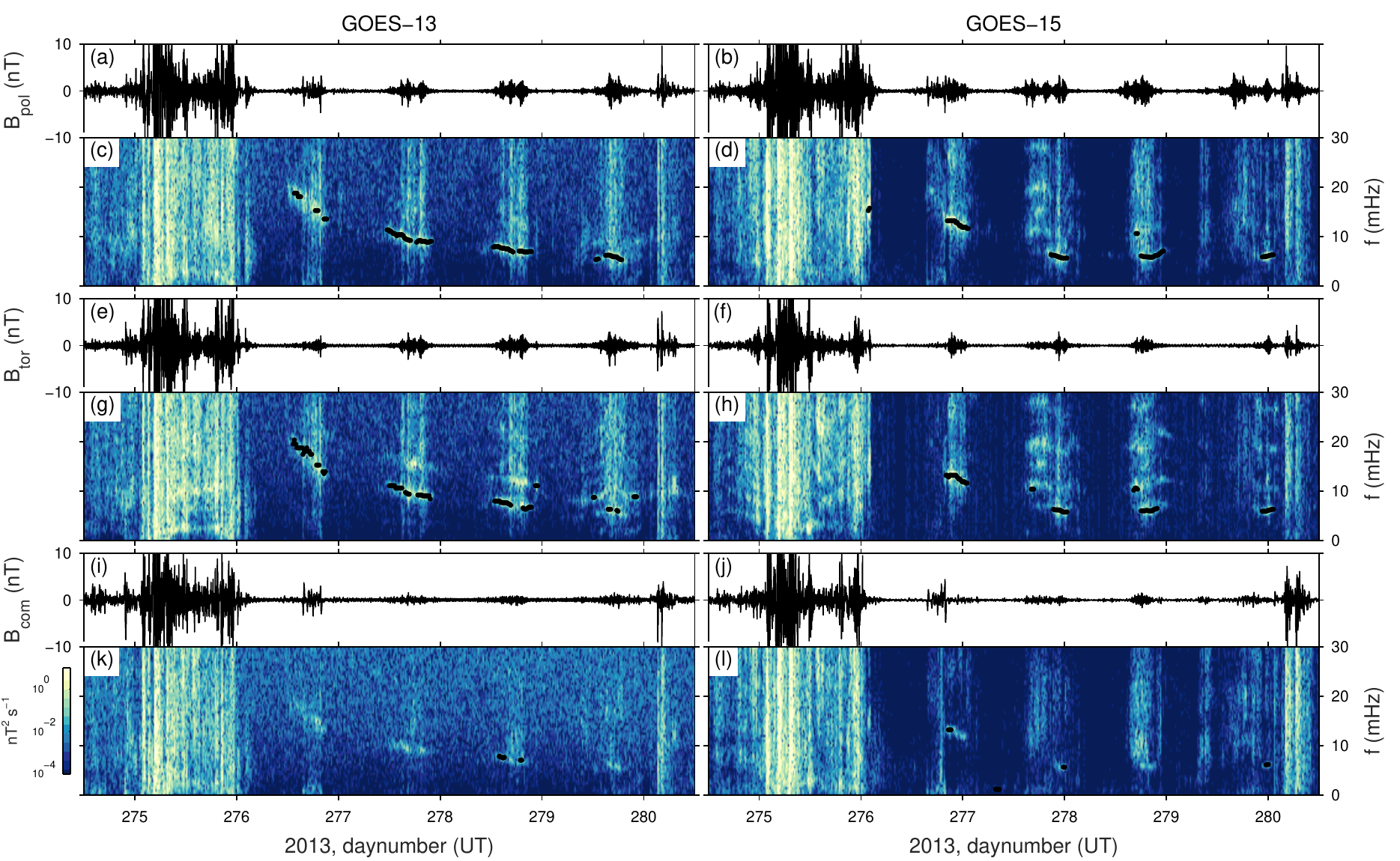}
\par\end{centering}
\caption{GOES-13 (left) and -15 (right) magnetometer observations of a long-lasting
ULF wave event displaying time-series of the radial/poloidal (a-b),
azimuthal/toroidal (e-f), and field-aligned/compressional (i-j) components
along with their respective whitened spectrograms (c-d, g-h, k-l).
Black dots indicate identified frequencies using an autocorrelation-based
method.\label{fig:spectrograms}}
\end{figure}

The case study event occured between 2-6 October 2013 (day of year
275\textendash 279). Figure~\ref{fig:spectrograms} shows both the
time series of each component of the magnetic field and their corresponding
spectrograms (using 1024-sample Hanning windows of the time-differenced
data with 50\% overlap). Compare this with the Audacity view of the
same event in Figure~\ref{fig:audacity} as well as the equivalent
audio, which can be found in the supporting information. The event
consisted of large-amplitude broadband waves lasting a day followed
by intervals of narrowband near-sinusoidal waves over the following
four days. The corresponding local times of these intervals, initially
identified by the citizen scientists through their use of the provided
spreadsheet, stayed relatively constant throughout at $\sim$07\textendash 17h,
thus their duration each day in the data is a spatial and not temporal
effect with both spacecraft encountering the waves each day/orbit.
Oscillations were predominantly in the poloidal component (a feature
noted by the citizen scientists) which had 50\% of the total power
overall, followed by the toroidal (28\%) and compressional (22\%)
components. These percentages varied by $\pm15\%$ when investigating
subintervals.

It is apparent from the spectrograms that the frequencies of the narrowband
waves decreased throughout the event, spanning the Pc3\textendash 5
frequency bands. It should be noted that this feature was what initially
alerted the citizen scientists to the event, as it was far more striking
through listening to the audio than cursorily glancing at spectrograms
alone. Therefore, the event was discovered thanks to the combined
sonification and exploratory citizen science elements of MUSICS. The
citizen scientists performed an initial characterisation of the frequency
decrease with time by using Audacity's spectral tools (e.g. see spectrogram
in Figure~\ref{fig:audacity}), converting frequencies from the audio
to their physical values in the provided spreadsheet. Here we perform
a more thorough frequency-time analysis. Because of the large variance
associated with the spectrogram's spectral estimators we opt to quantify
the frequencies using an autocorrelation method based on \citet{tolonen00},
widely used in audio analysis and described in Appendix~\ref{subsec:Pitch-Detection}.
The corresponding frequencies are indicated in Figure~\ref{fig:spectrograms}
as the black dots, which overall show good agreement with the lowest
harmonics of the narrowband waves in the spectrograms. There are a
few windows at the beginning or end of some days' observed waves where
a higher harmonic seems to have been selected due to little power
at lower frequencies. These instances have been manually removed in
further analysis.

\begin{figure}
\begin{centering}
\includegraphics[width=1\columnwidth]{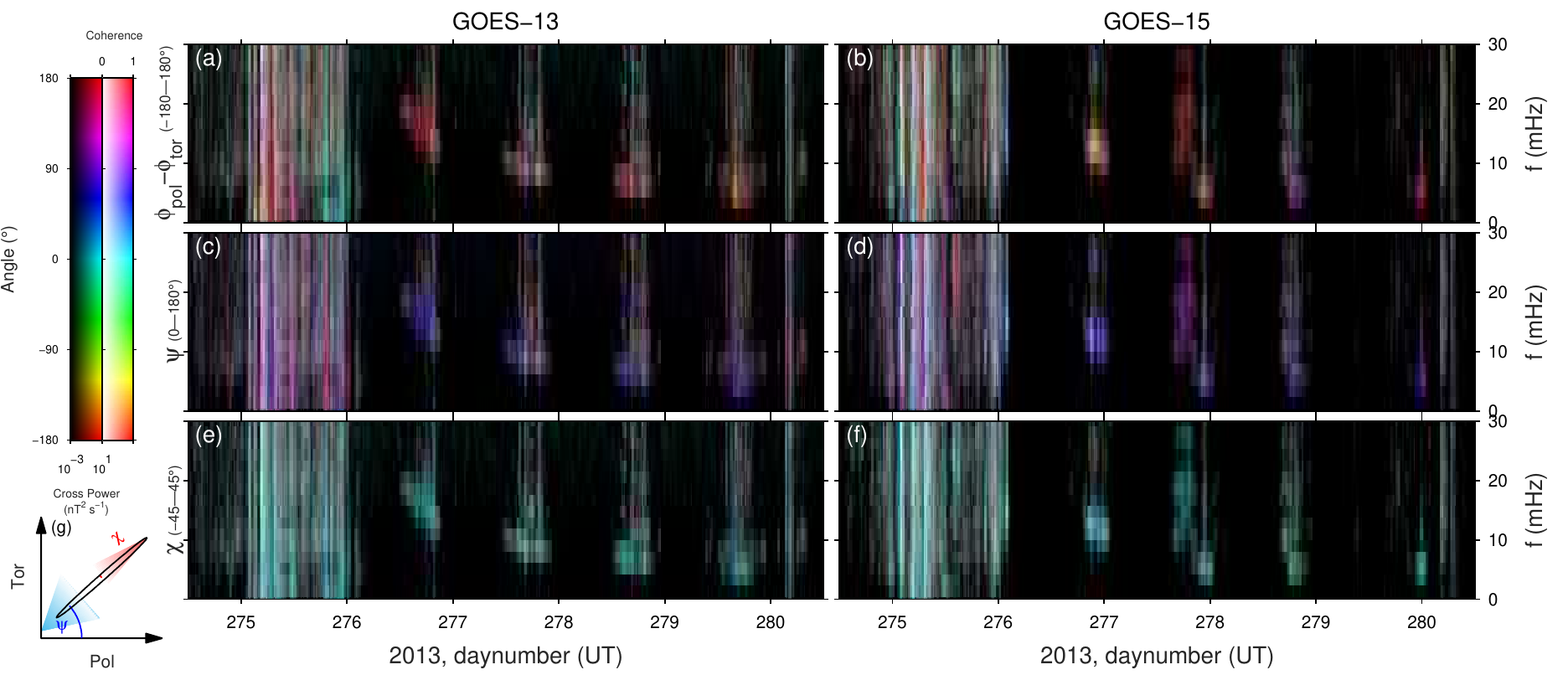}
\par\end{centering}
\caption{Polarisation parameters displayed using the HSV colour model depicting
angles (hue), coherence (saturation), and cross-power (value). Panels
correpond to (a-b) the phase difference between poloidal and toroidal
components, (c-d) the orientation angle of the polarisation ellipse,
(e-f) and the ellipticity angle. The latter two angles are depicted
in panel (g) along with the average polarisation ellipse (black) and
angular spreads (coloured areas).\label{fig:polarisation}}
\end{figure}

To ascertain the polarisation properties of the waves, the spectral
matrix of each 1024-sample window of differenced data was calculated
for all magnetic field component pairs using Welch\textquoteright s
overlapped averaged periodogram method with 8 segments (see Appendix~\ref{subsec:Welch}).
The cross phase between the poloidal and toroidal components ($\phi_{pol}-\phi_{tor}$)
are shown in Figure~\ref{fig:polarisation}a-b as the hue, revealing
they were largely close to antiphase with one another though with
considerable variability. To better quantify this the average phase
difference, weighted by both cross-power and coherence, and its spread
were calculated. These were $\phi_{pol}-\phi_{tor}=-173\pm52^{\circ}$,
$\phi_{pol}-\phi_{com}=-16\pm55^{\circ}$, and $\phi_{tor}-\phi_{com}=-179\pm45^{\circ}$.
The parameters of the polarisation ellipse in the transverse plane
were also calculated \citep{arthur76}, namely the orientation $\psi$
and ellipticity $\chi$ angles whose definitions are depicted in Figure~\ref{fig:polarisation}g.
The ellipse's orientation (Figure~\ref{fig:polarisation}c-d) shows
no systematic change with local time across the four wave intervals
with a weighted average value and spread of $\psi=42\pm30^{\circ}$.
Similarly there is little change in the ellipticity angle (Figure~\ref{fig:polarisation}e-f)
and its value is very low, i.e. close to plane polarisation, at $\chi=3\pm12^{\circ}$.
The average polarisation ellipse of the event is depicted in Figure~\ref{fig:polarisation}g.

Similar cross phase analysis was performed between the two GOES spacecraft
(not shown) in order to estimate the waves' azimuthal wavenumbers
$m$. However, the coherence was found to be poor ($<0.3$) throughout
and thus a reliable cross phase could not be determined. Given the
azimuthal separation of the two spacecraft of $60^{\circ}$, corresponding
to a maximum determinable $m$ of 6, such low coherence may indicate
higher wavenumbers.

It should be noted that a search for ground magnetometer signatures
of this event throughout the IMAGE network yielded no waves matching
the frequencies observed by GOES. This may also point towards high
$m$ waves since these get screened by the ionosphere making them
difficult to detect on the ground \citep{hughes76a}.

\subsection{Analysis}

\begin{figure}
\begin{centering}
\includegraphics{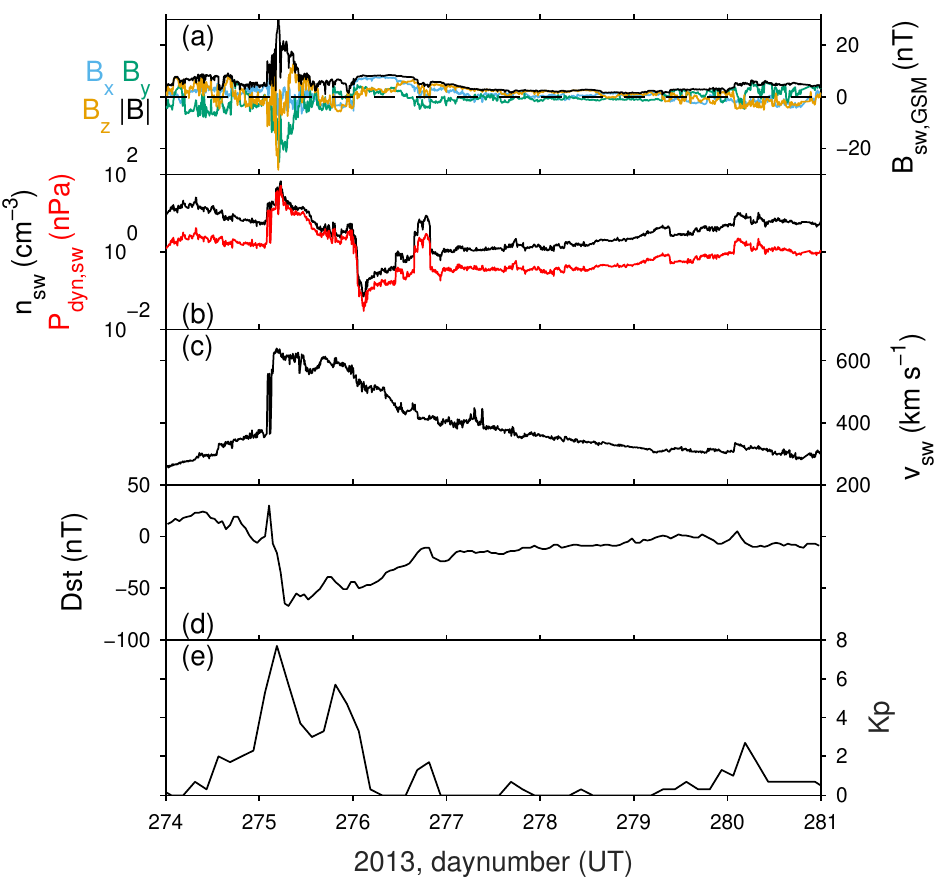}
\par\end{centering}
\caption{Solar wind and geomagnetic conditions showing (a) the IMF strength
(black) and GSM-components (x, y, z as blue, green, orange), (b) solar
wind density (black) and dynamic pressure (red), (c) solar wind speed,
and the $Dst$ (d) and $Kp$ (e) geomagnetic activity indices.\label{fig:solar-wind}}
\end{figure}

The citizen scientists looked into solar wind observations throughout
the event, taken from the OMNI database, revealing a Coronal Mass
Ejection (CME) as shown in Figure~\ref{fig:solar-wind}. The start
of the initial broadband wave activity corresponded with the arrival
of the CME shock, indicated by the sharp increase in the solar wind
density and speed (panels b and c respectively) as well as interplanetary
magnetic field (IMF) strength (panel a), with these waves lasting
for the duration of the CME's subsequent sheath region. The large
increase in the solar wind dynamic pressure associated with the CME
would have moved the magnetopause standoff distance to within geostationary
orbit based on the \citet{shue98} model. However, from the positions
of the two GOES spacecraft with time, they should not have encountered
the magnetopause. Indeed there was no indication from the original
magnetometer data of any magnetopause crossings.

Following the CME sheath was a magnetic cloud, a region of low density
and elevated IMF strength (compared to background values) along with
slow rotations in IMF orientation. The narrowband waves, which were
first very briefly observed by GOES-15 at the start of day 276 , seemed
to commence around the time of the arrival of the magnetic cloud,
though the waves lasted beyond the cloud's duration. We note that
while the start of these waves could clearly be heard in the audio
and was identified by the autocorrelation algorithm (see Figure~\ref{fig:spectrograms}d),
this was not clear at all from looking at the time series or spectrogram.
To check for the possibility that the waves were directly driven by
the solar wind we looked at 3~s resolution plasma and magnetic field
data from the ARTEMIS spacecraft in lunar orbit \citep{auster08,mcfadden08a,angelopoulos10}
but found no similar oscillations present.

The CME's arrival at the magnetosphere triggered a geomagnetic storm,
as indicated by the $Dst$ and $Kp$ activity indices (Figure~\ref{fig:solar-wind}d-e),
which was the largest of 2013 as measured by $Kp$ and the 10th largest
by $Dst$. The occurrence of the narrowband waves coincided with the
recovery phase of the storm.

\begin{figure}
\begin{centering}
\includegraphics[width=1\columnwidth]{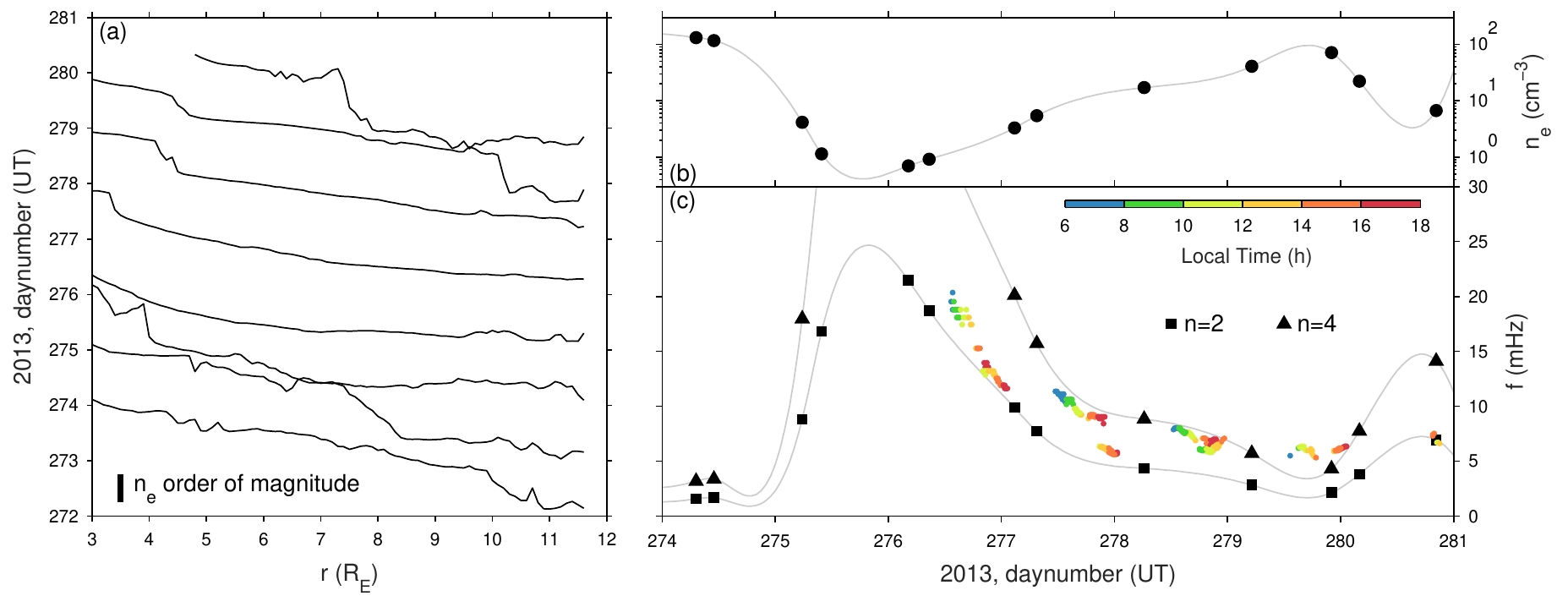}
\par\end{centering}
\caption{(a) THEMIS magnetospheric radial plasma density profiles at 19h LT
stacked by time. (b) Plasma density at geostationary orbit. (c) Estimated
frequencies of even mode poloidal field line resonances based on the
density observations. The observed frequencies are overlaid and coloured
by local time.\label{fig:density}}

\end{figure}

During the event the THEMIS \citep{angelopoulos08} spacecraft had
apogees in the dusk sector, with their outbound trajectories crossing
geostationary orbit at around 19h local time. This unfortunately meant
that they did not observe the narrowband ULF waves and their radial
extent. Using spacecraft potential inferred density measurements \citep{bonnell08,mcfadden08b}
from outbound crossings of THD and THE, we extended the electron density
radial profile survey of \citet{archer15b}. The evolution of these
profiles are depicted in Figure~\ref{fig:density}a. This reveals
an extended plasmasphere before the event, with the CME and resulting
storm causing the erosion of magnetospheric plasma and recession of
the plasmapause. Refilling of magnetospheric plasma occurred throughout
the time of the narrowband waves. This is shown more clearly for the
electron density measurements at geostationary orbit displayed in
Figure~\ref{fig:density}b. The apparent refilling rate approximately
followed a two-step process with exponential growth rates of

\begin{equation}
\frac{d\log_{10}\left(n_{e}\right)}{dt}=\begin{array}{cc}
0.77\pm0.05\,\mathrm{day^{-1}} & \mathrm{Early}\\
0.46\pm0.07\,\mathrm{day^{-1}} & \mathrm{Late}
\end{array}
\end{equation}
corresponding to $\sim1-10\,\mathrm{cm}^{-3}\,\mathrm{day^{-1}}$
initially, followed by $\sim13-33\,\mathrm{cm}^{-3}\,\mathrm{day^{-1}}$.
These are consistent with previously reported observations \citep{lawrence99}.
It is natural to think that the narrowband waves were field line resonances
(FLRs) whose decreasing frequency was a result of the increasing magnetospheric
density and thus decreasing Alfv\'{e}n speed.

The average frequency of poloidal field line resonances at geostationary
orbit across the dayside were estimated by solving the \citet{singer91}
equation applied to a T96 magnetic field model \citep{t95,t96}, using
the median conditions over the event combined with the THEMIS density
observations. An average ion mass of 7.5~amu \citep{sandhu16} was
used along with the usual power law form for the density along the
field lines using exponent $\alpha=1$ \citep{denton15}. Changing
the fixed parameters within their respective interquartile ranges
affected results by less than 10\%. The estimated second and fourth
harmonics are shown in Figure~\ref{fig:density}c, since even modes
should be most prominent in magnetic field data close to the magnetic
equator. While local time differences at geostationary orbit from
T96 alone are negligible, these calculations assume that the density
measured near dusk are representative of all dayside local times,
which is not true in general. However, Figure~\ref{fig:density}c
also shows the identified frequencies of the waves from both spacecraft
coloured by local time, revealing several instances of similar frequencies
observed simultaneously at different local times and a near continuity
of frequencies with time. This suggests a global expansion of the
plasmasphere throughout the event. There is good agreement between
the estimated frequencies and those observed by GOES, thus the refilling
can explain the decreasing frequency of the observed waves across
the event. Interestingly as the event progresses there appears to
be a change in the FLR harmonic as the frequency drops, seemingly
going from predominantly the second to later the fourth harmonic.

It is known that field line resonant frequencies at geostationary
orbit exhibit dependence on solar wind and geomagnetic conditions
\citep[e.g.][]{takahashi10}. Hourly averages of the observed frequency
and conditions across this event were taken, with the correlation
coefficients and their standard errors displayed in Table~\ref{tab:correlations}
(rank order correlation coefficients were similar). This was also
done for the second harmonic FLR frequency $f_{2}$ where instances
of what appears to be the fourth harmonic were corrected by the ratio
of the second and fourth harmonics from our FLR frequency estimates.
In most cases $f_{2}$ shows ever so slightly higher levels of correlation.
The correlation between estimated FLR frequencies with hourly averaged
conditions from the \citet{archer15b,archer17} survey across half
a solar cycle are also shown in the dawn, noon and dusk sectors for
comparison, where the standard errors are less than $\pm0.03$. As
pointed out by the citizen scientists, for this event the frequency
was highly correlated to the solar wind speed and IMF strength. The
latter was predominantly due to the GSM z component, despite the IMF
being northward throughout the intervals of narrowband waves, with
a fair level of correlation also due to the x component. These results
may have been down to chance for this specific event, given there
is little to no correlation between geostationary FLR frequencies
and the IMF in general. While around noon and dusk it is known that
there is some degree of correlation between the solar wind speed and
geostationary FLR frequencies, the observed correlation for this event
is much higher. Similar levels of correlation are typically expected
for the $Dst$ index though the correlation for this event, while
fair, weren't as strong. This highlights the potential importance
of the storm phase, and thus the time-history of the magnetosphere,
in density and FLR frequency models.

\begin{table}
\resizebox{\textwidth}{!}{%
\begin{tabular}{c|ccccccccc}
\multicolumn{2}{c}{} & $B_{sw}$ & $B_{x,sw}$ & $B_{y,sw}$ & $B_{z,sw}$ & $n_{sw}$ & $v_{sw}$ & $Dst$ & $AE$\tabularnewline
\cline{3-10} 
\multicolumn{2}{c}{$f$} & $0.92\pm0.03$ & $0.72\pm0.08$ & $-0.27\pm0.16$ & $0.80\pm0.06$ & $0.28\pm0.16$ & $0.92\pm0.03$ & $-0.71\pm0.09$ & $0.00\pm0.17$\tabularnewline
\multicolumn{2}{c}{$f_{2}$} & $0.96\pm0.01$ & $0.70\pm0.09$ & $-0.34\pm0.15$ & $0.84\pm0.05$ & $0.30\pm0.16$ & $0.94\pm0.02$ & $-0.70\pm0.09$ & $-0.23\pm0.16$\tabularnewline
\multirow{3}{*}{\citet{archer15b,archer17}} & Dawn & 0.04 & 0.02 & -0.07 & -0.06 & -0.1 & 0.24 & -0.17 & 0.14\tabularnewline
 & Noon & 0.02 & -0.05 & -0.04 & 0.01 & -0.21 & 0.54 & -0.41 & 0.17\tabularnewline
 & Dusk & 0.17 & -0.01 & 0.01 & -0.06 & -0.20 & 0.52 & -0.54 & 0.17\tabularnewline
\end{tabular}}

\caption{Correlation coefficients of hourly-averaged frequencies with solar
wind and geomagnetic conditions.\label{tab:correlations}}

\end{table}

\subsection{Discussion}

First we discuss the possible driving mechanisms of the observed narrowband
waves in this event. We have already excluded the possibility that
these waves were directly driven by the solar wind since similar oscillations
were not present upstream in particle or magnetic field measurements.
However, significant changes in the solar wind dynamic pressure were
present throughout and perhaps these could have resonantly excited
the observed field line resonances. It is expected that waves generated
by this mechanism are toroidally polarised and have low $m$ \citep{tamao65,southwood74,allan86,mann98},
unlike the observed event. Similarly, Alfv\'{e}n waves excited via
the Kelvin-Helmholtz instability would be expected to also have low
$m$ and to exhibit a reversal in polarisation about noon due to the
change in the flow direction within the magnetosheath around the subsolar
point \citep{dungey70,samson71,agapitov09}, however, no such reversal
was seen. Indeed high $m$ waves, as we suppose is the case for this
event, are a signature of internally and not externally driven waves
\citep{southwood69,le17}. Furthermore, the even harmonic field line
resonances observed imply an energy source which acts asymmetrically
about the equator, whereas externally driven waves tend to be more
often symmetric about this plane.

Therefore, we conclude that the processes generating the narrowband
waves likely were internal to the magnetosphere. Based on typical
periodicities, energy may be transfered to waves via drift and/or
bounce resonances with radiation belt electrons, ring current ions
or the background ion population. It has been suggested that observed
poloidal waves may be excited by bounce resonance with energetic (predominantly
H+) ions and that these are likely second harmonic FLRs \citep{southwood69,hughes78,glassmeier99}.
\citet{ferradas18} showed using Van Allen Probes observations that
this particular geomagnetic storm injected ions at energies above
$\sim10\,\mathrm{keV}$. This therefore demonstrates that there was
free energy available for conversion to wave power. We calculate the
pitch-angle averaged minimum bounce frequencies of these ions based
on their lower energies to be approximately 2~mHz (O+), 4~mHz (He+)
and 8~mHz (H+) respectively. These values can potentially explain
the change in harmonic throughout the event. While the eigenfrequencies
of the field lines were decreasing throughout the event due to the
plasmasphere refilling, the bounce frequencies are unaffected by this.
Therefore, as the frequency of the second harmonic dropped to around
the lowest possible bounce frequency, the ion bounce resonance became
more effective at driving fourth harmonic waves. This hypothesis,
coupled with the lack of change in polarisation from poloidal to toroidal
mode Alfv\'{e}n waves suggests that continuous driving occurred throughout
the event.

This event shares some characteristics of a type of ULF wave known
as giant pulsations: highly monochromatic poloidal oscillations in
the Pc4 (7\textendash 22~mHz) band that are localised in latitude,
have large azimuthal wavenumbers, and can continue for two or more
days \citep[e.g.][]{rostoker79,green85}. They are thought to be caused
by protons with energies of 5\textendash 30 keV drifting from the
nightside. However, giant pulsations are fundamental mode FLRs observed
on the ground concentrated in the dawn sector (though they have recently
been shown to also occur at dusk \citep{motoba15}) which occur during
quiet or late storm recovery periods and most often at solar minimum.
These properties are all unlike this event therefore it cannot be
deemed to be a giant pulsation.

Other long-duration Pc5 waves have been published though these share
little similarities with this event since they tend to be predominantly
compressional waves in the nighttime sector during the start of storm
times \citep{takahashi85,takahashi87a,takahashi87b}. To our knowledge
only two truly similar events to this have been previously reported.
\citet{sarris09a} presented an apparently rare narrowband Pc5 event
in GOES magnetometer data from 1997 lasting five days during the recovery
phase of a storm. The waves were chiefly poloidally oriented (though
packets of phase mixing from poloidal to toroidal mode were observed
within the event \citep{sarris09b}) with frequencies decreasing with
time from 9~mHz to 5~mHz over the course of the event. Density measurements
from the LANL spacecraft, also in geostationary orbit, revealed local
evidence of a plasmasphere refilling process associated with the observed
slow frequency decrease. The waves' azimuthal extent shrank throughout
the course of the event, initially spanning some 12~h across the
dayside and by the end of the event being concentrated over 7~h of
local time in the noon and post-noon sectors. This feature is unlike
our observations, where the azimuthal extent remained widespread and
almost constant throughout. Also unlike the event presented here,
\citet{sarris09a} found signatures of the waves in ground magnetometer
data revealing large $m$ values between 20 and 55 and an eastward
propagation in all local time sectors. They also concluded that these
waves were likely driven by internal particle anisotropies and not
via external mechanisms. \citet{korotova16} reported a Pc4 event
observed by Van Allen Probes, THEMIS and GOES over 1.5~days during
the late recovery phase of a moderate storm in 2014. These were observed
throughout most of the morning sector except for by GOES which only
observed the waves in the late afternoon sector, unlike the event
presented here. They showed that these were even harmonic (likely
the second harmonic based on previously reported frequencies) poloidal
field line resonances and that they spanned a number of L-shells,
dissimilar to the \citet{sarris09a,sarris09b} event. The frequencies
of these waves decreased over the course of successive orbits from
17~mHz to 12~mHz based on the GOES observations and down to $\sim8\,\mathrm{mHz}$
at geostationary L-shells based on the other spacecreaft measurements,
with some evidence of plasmasphere refilling occurring. Again they
attributed the source of the pulsations as due to proton bounce resonance.
Similar events to the case study are therefore little discussed in
the literature, and indeed thought to be rare. The MUSICS project
was thus not designed with this class of event in mind, nonetheless
the combination of sonification and exploratory citizen science uncovered
one such example.

\begin{figure}
\begin{centering}
\includegraphics{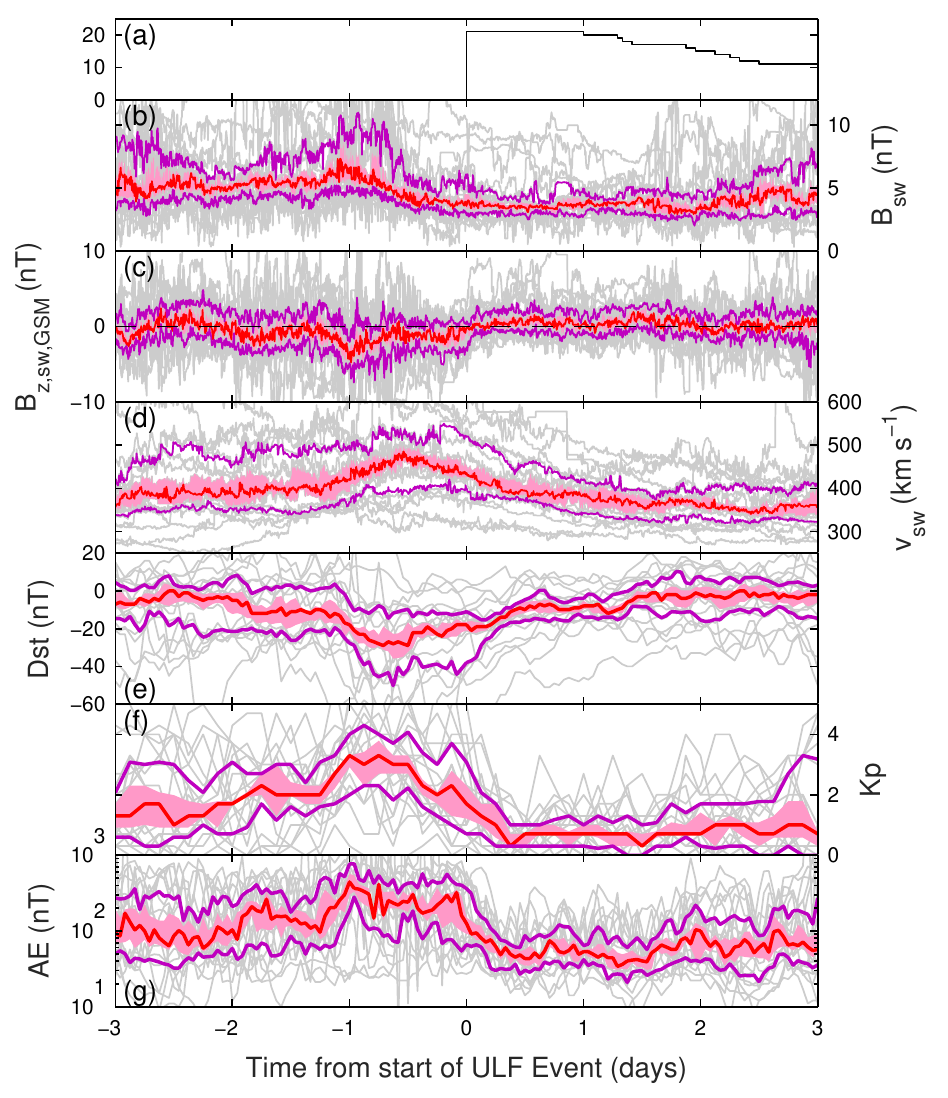}
\par\end{centering}
\caption{Superposed epoch analysis of the solar wind and geomagnetic conditions
during the 21 identified ULF wave events. Displayed are the IMF strenth
(b), IMF GSM-z component (c), solar wind speed (d), and the $Dst$
(e), $Kp$ (f) and $AE$ (g) geomagnetic activity indices. All time-series
are shown in grey with the medians (red) and corresponding standard
error (red shaded) as well as interquarile ranges (purple) also displayed.
Panel (a) indicates the number of ULF wave events occurring at each
epoch time.\label{fig:superposed-epoch}}

\end{figure}

To better put this event into context and to test whether such events
are indeed rare, we navigated the sonified GOES-13 poloidal component
data searching for similar events thereby taking advantage of the
reduced timeframe of the sonified data. More precisely this search
looked for decreasing frequency narrowband waves spanning at least
one day, revealing 21 events in 2013 alone. We note that many of these
events were much more subtle, even in the sonified data, than the
case study being often weaker and/or superposed with other waves.
Only through the case study having been identified by the citizen
scientists and through auditory training (listening to the event numerous
times) was it possible to then pore through the audio to easily identify
further events \citep[cf.][]{whitton14}. The start of each event
was identified as the beginning of the first interval (where each
interval as per the case study is a fraction of a day in duration
due to the limited local times of the waves) of the decreasing-frequency
narrowband poloidal waves, with the duration of the event being the
time until the end of the very last interval of narrowband waves which
continued the decreasing-frequency trend. No overlap in events occurred.
Figure~\ref{fig:superposed-epoch}a shows the number of events in
occurrence as a function of time from their start. The median duration
was 3.0~days and the longest event lasted 5.4~days, with the total
duration of all events being 60.5~days i.e. 17\% of the entire year.
Therefore, such events cannot be considered rare.

A superposed epoch analysis of the solar wind and geomagnetic conditions
for these 21 events was performed to ascertain the typical driving
conditions, which are shown in Figure~\ref{fig:superposed-epoch}b-g
where the start of the ULF wave event was set as the zero epoch and
the median and interquartile ranges (and their respective standard
errors) were calculated at each epoch time. Of course, since the ULF
wave events have a limited extent within local time an ambiguity between
temporal and spatial effects exists within this analysis. The result
will be a smearing out of the superposed epoch analysis over timescales
of at most one day. The clearest and most significant results are
in the indices of geomagnetic activity revealing, like with the case
study event, increased activity for a period of about a day immediately
before the start of the wave events, indicative of storm conditions
beforehand. All quartiles of the $Dst$ index show slow returns to
pre-storm levels over the course of several days starting around the
time of the events. Furthermore there was a one-to-one correspondence
between many of these events and geomagnetic storms listed at \href{http://www.spaceweatherlive.com}{SpaceWeatherLive.com}.
Thus such ULF wave events typically occur (at least initially) during
the recovery phase of geomagnetic storms. It is not clear, however,
whether all storms (meeting some criteria) result in these ULF wave
events. Storms are obviosuly more common during solar maximum, which
was the case for 2013, whereas the observations of \citet{sarris09a,sarris09b}
occurred during solar minimum, when storms occur less frequently and
thus such ULF wave events are likely less common.

While the superposed epoch signatures in activity were clear, they
proved less pronounced in the solar wind. All quartiles show some
enhancement in solar wind speed before the start of the event with
the speed returning to background levels over the course of a couple
of days. The enhanced speeds, however, are typically nowhere near
as pronounced as the case study. The IMF tends to exhibit a small
enhancement in magnitude and a tendency towards a slight southward
component about a day before the start of events. No significant variations
of the solar wind density with epoch time were found (not shown).
The superposed epoch analysis, while showing clear evidence of storms
beforehand, do not reveal clear overall signs of CMEs and/or magnetic
clouds which likely would remain even when considering the smearing
out effects of the analysis. Given the clear signatures in activity
(which would also be subject to the same smearing effect) but not
in the solar wind, this suggests that the waves are predominantlydriven
via internal processes and it is likely that many geoeffective drivers,
including for instance Corotating Interaction Regions \citep[e.g.][]{tsurutani11},
may also act as the causes of some of these ULF wave events. We leave
a more detailed individual analysis of each event to future work.

\section{Conclusions}

To potentially address some of the issues facing magnetospheric ULF
wave research due to the ``big data'' being produced, we introduce
sonification and citizen science. ULF waves were extracted from magnetometer
data at geostationary orbit and converted into audible sound. This
dramatically reduces the timeframe for analysis and takes advantage
of many of the benefits of the human auditory system compared to visual
analysis \citep{robinson56,hermann2002,divenyi05,oppenheim13,whitton14}.
We have shown that this sonified data can be used in exploratory citizen
science, an approach somewhat different from typical citizen science
projects.

The potential for the future use of sonification with citizen science
has been demonstrated through the first results discovered by a group
of London high school students. A case study event consisting of narrowband
waves of decreasing frequency spanning several days were found thanks
to the citizen scientists' aural exploration of the data. These corresponded
to second and fourth harmonic poloidal field line resonances across
the dayside magnetosphere. It was found that the event occurred during
the recovery phase of a CME-driven geomagnetic storm. Simultaneous
plasma density measurements revealed the initial erosion of magnetospheric
plasma upon arrival of the CME and the subsequent refilling process.
The identified frequencies of the observed waves agreed well with
estimates of the field lines' eigenfrequencies made by combining these
density measurements throughout this refilling process with a magnetic
field model, thereby explaining the frequencies throughout the event.
These frequencies highly correlated with the solar wind speed and
IMF strength to a degree much greater than usually expected, perhaps
highlighting the importance of the storm phase and the time-history
of the magnetosphere in the empirical modelling of magnetospheric
densities and FLR frequencies. We were able to rule out external driving
mechanisms such as direct or resonant driving of the waves from the
solar wind and excitation via the Kelvin-Helmholtz instability, concluding
that the waves were likely driven by an internal process such as the
ion bounce resonance \citep{southwood69,hughes78,glassmeier99}.

Events such as those presented in this manuscript are little discussed
in the literature and had previously been reported as rare \citep{sarris09a,sarris09b}.
However, by surveying the audio (and thus taking advantage of the
human auditory system's pattern recognition and blind source separation
abilities as well as the reduced timeframe of the sonified data) we
found this not to be the case. During 2013 alone 21 events were discovered
and superposed epoch analysis revealed that these tend to occur following
storms, though no clear solar wind signature emerged. It may be possible
that several different solar wind transients are capable of trigerring
such events. Further work is required to understand the range of driving
conditions and properties of all of these events.

In conclusion, by combining sonification with citizen science it is
possible to arrive at unexpected research results - the science results
presented in this paper were possible only because of citizen scientists'
unprescribed aural exploration of ULF wave data. Indeed, the MUSICS
project was not set up with the particular class of ULF wave event
presented in mind from the outset. Therefore this approach has the
potential to address some of the current challenges facing current
magnetospheric ULF wave research. Sonified data could also be used
within the more standard citizen science approach in the future, once
a well-defined set of ULF wave event classifications and prescribed
tasks for citizen scientists have been established. Given the potential benefits, the audible GOES ULF wave dataset will now be made publicly available by NOAA. Finally, we hope that similar efforts become more widespread in the future and plan
to implement them further across different datasets.
\begin{acknowledgments}
M.O. Archer thanks the students of Eltham Hill School who contributed
to this study: Megan Wilkin, Hanan Mohamed, Isobel Currie, David Perry
and their teacher Russel Dobson. M.D. Hartinger was supported by NASA
NNX17AD35G. We thank NOAA's National Centers for Environmental Information
for access to GOES magnetometer data at \href{http://www.ngdc.noaa.gov/stp/satellite/goes/}{http://www.ngdc.noaa.gov/stp/satellite/goes/}. We acknowledge NASA contract
NAS5-02099 and the THEMIS Mission, specifically J. W. Bonnell and
F. S. Mozer for EFI data; C. W. Carlson and J. P. McFadden for ESA
data; and K. H. Glassmeier, U. Auster and W. Baumjohann for the use
of FGM data provided under the lead of the Technical University of
Braunschweig and with financial support through the German Ministry
for Economy and Technology and the German Center for Aviation and
Space (DLR) under contract 50 OC 0302. THEMIS data and analysis software (SPEDAS) are available at \href{http://themis.ssl.berkeley.edu}{http://themis.ssl.berkeley.edu}. The OMNI data was obtained
from the NASA/GSFC OMNIWeb interface at \href{http://omniweb.gsfc.nasa.gov}{http://omniweb.gsfc.nasa.gov}.
We thank the institutes who maintain the IMAGE Magnetometer Array:
Troms{\o} Geophysical Observatory of UiT the Arctic University of
Norway (Norway), Finnish Meteorological Institute (Finland), Institute
of Geophysics Polish Academy of Sciences (Poland), GFZ German Research
Centre for Geosciences (Germany), Geological Survey of Sweden (Sweden),
Swedish Institute of Space Physics (Sweden), Sodankyl\"{a} Geophysical
Observatory of the University of Oulu (Finland), and Polar Geophysical
Institute (Russia).
\end{acknowledgments}

\appendix

\section{Spectral Methods}

Here we detail some of the spectral methods used which are well used
in the fields of audio analysis but not widely applied in space physics.

\subsection{Pitch Detection\label{subsec:Pitch-Detection}}

The autocorrelation of any periodic function is also a periodic function,
exhibiting positive peak autocorrelation values of unity at lags given
by integer multiples of the underlying period. Due to the finite length
of real time-series as well as the typical application of windowing
functions, however, the height of these peaks in reality will decrease
as the lag increases, reaching zero when this equals the length of
the window used. Here we present details of the autocorrelation pitch
detection method used, which was based on that of \citet{tolonen00}
implemented within Audacity.

Autocorrelation functions were computed for 1024-point Hanning windows
of the (non-differenced) magnetic field data and we limit ourselves
to positive lags only. Half-wave rectification was performed, whereby
any negative autocorrelation values were set to zero thereby leaving
only a series of positive peaks in the case of a periodic signal.
To remove repeated peaks originating from the same underlying periodicity,
integer factor time-scaled copies of these autocorrelations were subtracted
from the original. This was followed by again clipping to positive
values only each time. This subtraction procedure was repeated for
all prime numbers up to 11. The resulting function, named the Enhanced
Summary Autocorrelation Function (ESACF) by \citet{tolonen00}, should
leave positive peaks only at the fundamental periodicities of the
signal.

Given the values of the ESACF peaks are not as simple to interpret
as a standard autocorrelation, we specify two criteria for selecting
significant peaks for each window of the data. Firstly, we calculated
the envelope function of the autocorrelation due to the Hanning window
used. Our threshold for peaks in ESACF was chosen to be at least half
that expected for perfect correlation. Secondly, we used a bootstrapping
technique \citep{efron93}, generating 25,000 realisations of 1024-sample
uncorrelated gaussian white noise. The same windowing function was
again applied and the autocorrelations calculated for each realisation.
At each lag, quantiles of the autocorrelations were computed to give
a confidence interval for uncorrelated gaussian white noise. The $1-\alpha$
quantile at each lag corresponds to the local confidence level at
the desired significance $\alpha$. However, we are performing multiple
comparisons corresponding to the search for peaks present at any value
of lag. Therefore, the global confidence level must be used to maintain
the desired significance. We use the Bonferroni correction, which
sets the confidence level to $1-\frac{\alpha}{N}$ where $N$ is the
number of comparisons i.e. the length of the window in this case \citep{bonferroni,dunn58,dunn61}.
We then required that the prominence (or intrinsic height) of peaks
in ESACF be greater than the global 95\% confidence level for uncorrelated
gaussian white noise.

\subsection{Welch's Method\label{subsec:Welch}}

Welch\textquoteright s overlapped averaged periodogram method \citep{welch67,stoica05}
is a way of estimating the power spectral matrix of a signal, reducing
noise at the expense of frequency resolution. Each 1024-sample set
of data was split up into 8 different data segments with 50\% overlap.
Hanning windows are then applied to each segment and the Fast Fourier
Transform $\mathcal{F}\left(x_{i}\left[t\right]\right)$ computed
for the windowed segments, where $x_{i}\left[t\right]$ represent
the segments of the original time-series $x\left[t\right]$. In the
case of autopower spectra, the periodograms for each segment are averaged
together, which we denote by angular brackets with a subscript index
$\left\langle \mathcal{F}\left(x_{i}\left[t\right]\right)\mathcal{F}\left(x_{i}\left[t\right]\right)^{\star}\right\rangle _{i}$,
reducing the variance of the individual power measurements. For cross
spectra between $x\left[t\right]$ and $y\left[t\right]$, the calculation
is $\left\langle \mathcal{F}\left(x_{i}\left[t\right]\right)\mathcal{F}\left(y_{i}\left[t\right]\right)^{\star}\right\rangle _{i}$.
Finally, the magnitude squared coherence using this method is given
by
\begin{equation}
C_{xy}=\frac{\left|\left\langle \mathcal{F}\left(x_{i}\left[t\right]\right)\mathcal{F}\left(y_{i}\left[t\right]\right)^{\star}\right\rangle _{i}\right|^{2}}{\left\langle \mathcal{F}\left(x_{j}\left[t\right]\right)\mathcal{F}\left(x_{j}\left[t\right]\right)^{\star}\right\rangle _{j}\left\langle \mathcal{F}\left(y_{k}\left[t\right]\right)\mathcal{F}\left(y_{k}\left[t\right]\right)^{\star}\right\rangle _{k}}
\end{equation}

\end{document}